\documentclass{ws-ijns}
\pdfoutput=1
\usepackage{graphicx}
\usepackage[english]{babel}
\usepackage[super,sort,compress]{cite}
\usepackage{textcomp}
\usepackage{float}
\begin{document}

\title{INTEGRATING EEG AND MEG SIGNALS TO IMPROVE MOTOR IMAGERY CLASSIFICATION IN BRAIN-COMPUTER INTERFACE}

\author{MARIE-CONSTANCE CORSI}
\address{Inria Paris, Aramis project-team, 75013, Paris, France\\
Sorbonne Universit\'es, UPMC Univ Paris 06, Inserm, CNRS, Institut du cerveau et la moelle (ICM) - H\^opital Piti\'e-Salp\^etri\`ere,\\ 
Boulevard de l\textquotesingle h\^opital, F-75013, Paris, France}

\author{MARIO CHAVEZ}
\address{CNRS UMR7225, H\^opital Piti\'e-Salp\^etri\`ere, Paris, France}

\author{DENIS SCHWARTZ, LAURENT HUGUEVILLE}
\address{Centre de NeuroImagerie de Recherche - CENIR,\\
Centre de Recherche de l'Institut du Cerveau et de la Moelle Epin\`ere, \\
Universit\'e Pierre et Marie Curie-Paris 6 UMR-S975, Inserm U975, CNRS UMR7225, Groupe Hospitalier Piti\'e-Salp\^etri\`ere, Paris, France}

\author{ANKIT N. KHAMBHATI}
\address{Department of Bioengineering, University of Pennsylvania, Philadelphia, PA, 19104, USA}

\author{DANIELLE S. BASSETT}
\address{Department of Bioengineering, University of Pennsylvania, Philadelphia, PA, 19104, USA \\ 
Department of Electrical and Systems Engineering, University of Pennsylvania, Philadelphia, PA, 19104, USA \\
Department of Physics, University of Pennsylvania, Philadelphia, PA, 19104, USA\\
 Department of Neurology, University of Pennsylvania, Philadelphia, PA, 19104, USA}

\author{FABRIZIO DE VICO FALLANI}
\address{Inria Paris, Aramis project-team, 75013, Paris, France\\
Sorbonne Universit\'es, UPMC Univ Paris 06, Inserm, CNRS, Institut du cerveau et la moelle (ICM) - H\^opital Piti\'e-Salp\^etri\`ere,\\ 
Boulevard de l\textquotesingle h\^opital, F-75013, Paris, France\\
E-mail: fabrizio.devicofallani@gmail.com}

\maketitle

\begin{abstract}
We adopted a fusion approach that combines features from simultaneously recorded electroencephalographic (EEG) and magnetoencephalographic (MEG) signals to improve classification performances in motor imagery-based brain-computer interfaces (BCIs). We applied our approach to a group of 15 healthy subjects and found a significant classification performance enhancement as compared to standard single-modality approaches in the alpha and beta bands. Taken together, our findings demonstrate the advantage of considering multimodal approaches as complementary tools for improving the impact of non-invasive BCIs.
\end{abstract}


\newpage
\begin{multicols}{2}

\section{Introduction}
Brain-computer interfaces (BCIs) exploit the ability of subjects to modulate their brain activity through intentional mental effort, such as in motor imagery (MI). BCIs are increasingly used for control and communication\cite{vidal_toward_1973, bozinovski_using_1988, wolpaw_braincomputer_2002,fifer_simultaneous_2014,carlson_brain-controlled_2013,lafleur_quadcopter_2013,jin_changing_2012,hwang_development_2012,kashihara_brain-computer_2014,naci_brains_2013, ortiz-rosario_brain-computer_2013}, and for the treatment of neurological disorders \cite{daly_braincomputer_2008,prasad_applying_2010,king_operation_2013,chatelle_brain-computer_2012,kim_real-time_2014, burns_brain-computer_2014}. 
 
 Despite their societal and clinical impact, many engineering challenges remain, from the optimization of the control features to the identification of the best mental strategy to code the user\textquotesingle s intent \cite{guger_brain-computer_2013}. Furthermore, between $15$ and $30~\%$ of the users are affected by a phenomenon called ``BCI illiteracy'' \cite{vidaurre_towards_2010} which consists in not being able to control properly a BCI even after several training sessions. BCI illiteracy particularly concerns MI-based BCIs because of the inherent difficulty to produce distinguishable brain activity patterns. \cite{allison_could_2010}. 
 
These challenges critically affect the usability of MI-based BCIs \cite{zickler_bci_2009} and have motivated, on the one hand, a deeper understanding of mechanisms associated with MI\cite{toppi_investigating_2014,kaiser_cortical_2014,perdikis_subject-oriented_2014,wander_distributed_2013}, and on the other hand the research of new features to enhance BCI performance for both healthy subjects and patients  \cite{vidaurre_co-adaptive_2011, pfurtscheller_hybrid_2010,pichiorri_sensorimotor_2011,pichiorri_brain-computer_2015}. In the latter case, hybrid and multimodal approaches adding respectively different type of biosignals \cite{pfurtscheller_hybrid_2010,muller-putz_towards_2015} and neuroimaging data, such as near-infrared spectroscopy (NIRS)\cite{sitaram_hemodynamic_2009,fazli_enhanced_2012,tomita_bimodal_2014,buccino_hybrid_2016} and functional magnetic resonance imaging (fMRI) \cite{perronnet_unimodal_2017}, have been proven to increase the overall performance.

Here, we consider magnetoencephalography (MEG), which carries complementary information in terms of source depth \cite{cuffin_comparison_1979} and conductivity\cite{geisler_surface_1961,delucchi_scalp_1962,cooper_comparison_1965,hamalainen_magnetoencephalography-theory_1993} sensitivities, but also radially/tangentially oriented dipole detection \cite{wood_electrical_1985,sharon_advantage_2007}. \cite{boto_new_2017} While previous studies have demonstrated the feasibility of BCI \cite{mellinger_meg-based_2007} and neurofeedback \cite{halme_comparing_2016}, based on MEG activity, the potential benefit of the combination with EEG  signals has been poorly explored. 
Indeed, such integration might have practical consequences in the light of the recent development of portable MEG sensors, based on optically pumped magnetometers \cite{boto_new_2017}. 

To address this gap in knowledge, we considered high-density EEG and MEG signals simultaneously recorded in a group of healthy subjects during a MI-based BCI task. We then propose a matching-score fusion approach to test the ability to improve the classification of motor-imagery associated with BCI performance.

\section{Materials and Methods}
\subsection{Simultaneous E/MEG recordings}
Fifteen~healthy subjects (aged $28.13\pm 4.10 $~years, 7~women), all right-handed, participated in the study. None presented with medical or psychological disorders. According to the declaration of Helsinki, written informed consent was obtained from subjects after explanation of the study, which was approved by the ethical committee CPP-IDF-VI of Paris. All participants received financial compensation at the end of their participation. 
MEG and EEG data were simultaneously recorded with, respectively, an Elekta Neuromag TRIUX\textsuperscript{\textregistered}~machine (which includes 204~planar gradiometers and 102~magnetometers) and with a $74$ EEG-channel system. The EEG electrodes positions on the scalp followed the standard 10-10 montage. EEG signals were referenced to mastoid signals, with the ground electrode located at the left scapula, and impedances were kept lower than 20~kOhms.  On average, 1.5~hours was needed for subjects preparation (i.e.  explaining the protocol, placing the electrodes, registering the EEG sensor positions and checking the impedances). M/EEG data were recorded in a magnetically shielded room with a sampling frequency of 1~kHz and a bandwidth of 0.01-300 Hz.  The subjects were seated in front of a screen at a distance of 90 cm. To ensure the stability of the position of the hands, the subjects laid their arms on a comfortable support, with palms facing upward. We also recorded electromyogram (EMG) signals from the left and right arm of subjects. Expert bioengineers visually inspected EMG activity to ensure that subjects were not moving their forearms during the recording sessions.
We carried out BCI sessions with EEG signals transmitted to the BCI2000 toolbox\cite{schalk_bci2000:_2004} via the Fieldtrip buffer\cite{oostenveld_fieldtrip:_2010}.

\subsection{BCI protocol}
We used the one-dimensional, two-target, right-justified box task \cite{wolpaw_wadsworth_2003}, where subjects had to perform a sustained MI (grasping) of the right hand to hit up-targets, while remaining at rest to hit down-targets. Each run consisted of 32~trials with up and down targets, consisting of a grey vertical bar displayed on the right portion of the screen, equally and randomly distributed across trials. \\
The experiment was divided into two phases: 
\begin{romanlist}[(ii)]
\item Training: The training phase consisted of five consecutive runs without any feedback. For a given trial, the first second corresponded to the inter-stimulus interval (ISI), where a black screen was presented to the subject. The target appeared and persisted on the screen during subsequent five seconds (from 1 s to 6 s). During this period subjects had to perform the instructed mental tasks. 
\item Testing: The testing phase consisted of six runs with a visual feedback. For a given trial, the first second corresponded to the ISI, while the target was presented throughout the subsequent five seconds, with the same modalities just as in the training phase. In the last three seconds (from 3 s to 6 s), subjects received a visual feedback to control an object that consists of a cursor (a ball here) that starts from the left-middle part of the screen and moves to the right part of the screen with fixed velocity. This gave a fixed of communication rate of $20$ commands/minute. Only vertical position was controlled by the subject's brain activity and it was updated every 28 ms. The aim is to hit the target with the ball according to the instructed mental tasks, i.e. MI for up-targets; resting for down-targets.
\end{romanlist}

\subsection{Signal processing and features extraction}
We considered both EEG and MEG activity, the latter consisting of magnetometer (MAG) and gradiometer (GRAD) signals which, given their physical properties, can be processed separately. \cite{hansen_meg:_2010} 

As a preliminary step, temporal Signal Space Separation (tSSS) \cite{taulu_spatiotemporal_2006} was performed using MaxFilter (Elekta Neuromag) to remove environmental noise from MEG activity. All signals were downsampled to 250~Hz and segmented into epochs of five seconds corresponding to the target period. To simulate online scenarios, no artifact removal method was applied. Expert bioengineers visually inspected the recorded traces to ensure that no major artifacts (e.g. MEG jumps, EEG pops) were present. After verification, we then kept all the available epochs. 

We computed for each sensor the power spectrum between 4 and 40~Hz, with a 1~Hz frequency bin resolution, for both MI and rest epochs. To this end, we used a multi-taper frequency transformation based on discrete prolate spheroidal sequences (Slepian sequences \cite{slepian_prolate_1978}) considered as tapers through the use of the Fieldtrip toolbox \cite{oostenveld_fieldtrip:_2010}. A $\pm$~0.5~Hz spectral smoothing through multi-tapering was applied.

At this stage, each epoch was characterized by a feature matrix $\textbf{M}_i$, containing the power spectrum values for every couple of sensor and frequency bin, and whose dimension was $74 \times 36$, $102 \times 36$ and $204 \times 36$, respectively for $i = EEG, MAG, GRAD$.

We adopted a semi-automatic procedure to extract the most relevant features from the matrices $\textbf{M}_i$ in the training phase.
First, we focused on sensors in the motor area contralateral to the movement (see Appendix \ref{appendix1}). In this way, the size of the feature matrices became $8\times 36$, $11\times 36$ and $22\times 36$ respectively for EEG, MAG, and GRAD.
Second, for each selected sensor and frequency bin, we performed a non-parametric cluster-based permutation t-test between the power spectrum values of the MI and rest epochs\cite{bozinovski_using_1988, bozinovski_controlling_2013}. To this end, we set a statistical threshold of $p<0.05$, false-discovery rate corrected for multiple comparisons, and 500~permutations. 

We finally extracted the $N_{f}$ most discriminant features within the standard frequency bands $b=~theta ~ (4-7~Hz),~alpha ~(8-13~Hz),~beta ~(14-29~Hz),~gamma ~(30-40~Hz)$.
This allowed us to identify, for each modality $i$ and band $b$, the best (sensor, frequency bin) couples to be used in the testing phase to compute the features  $\hat{m}$. 
Hence, the final feature vectors used for the classification are given by:
\begin{equation}
\zeta_{i,b}=[\hat{m}_{i,b}^1,\ldots,\hat{m}_{i,b}^{N_f}],
\end{equation} 
where $N_{f}=1\ldots,10$. 
The maximal limit of 10 was chosen based on the actual number of features (between 4 and 6) that we used in the recording sessions, conforming to the guidelines associated with similar MI-based BCI and EEG montages. \cite{schalk_bci2000:_2004}

\subsection{Classification, fusion, and performance evaluation}
We performed a separate classification for each value of $N_f$. Given the relatively small number of features, we used a five-fold cross-validation in a linear discriminant analysis-based (LDA) classification.\cite{fukunaga_introduction_1990, duda_pattern_2000} LDAs are particularly suited for two-class MI-based BCIs \cite{lotte_review_2007}. 

To integrate the information from different modalities we used a Bayesian fusion approach based on the weighted average method\cite{ruta_overview_2000,roli_analysis_2002,roli_multiple_2009}. Similar to what has been proposed for hybrid-BCI systems \cite{muller-putz_towards_2015}, we linearly combined the posterior probabilities $p_i$, obtained from the classification of each modality $i$, weighted by the parameter $\lambda_i $:
\begin{equation}
\lambda_i = \dfrac{p_{i}}{p_{EEG}+p_{MAG}+p_{GRAD}}.
\end{equation}
In this manner, a higher weight was assigned to the modality that best classified the data (see Figure \ref{fig1}).
 
\begin{figurehere}
\begin{center}
\includegraphics[scale=0.24]{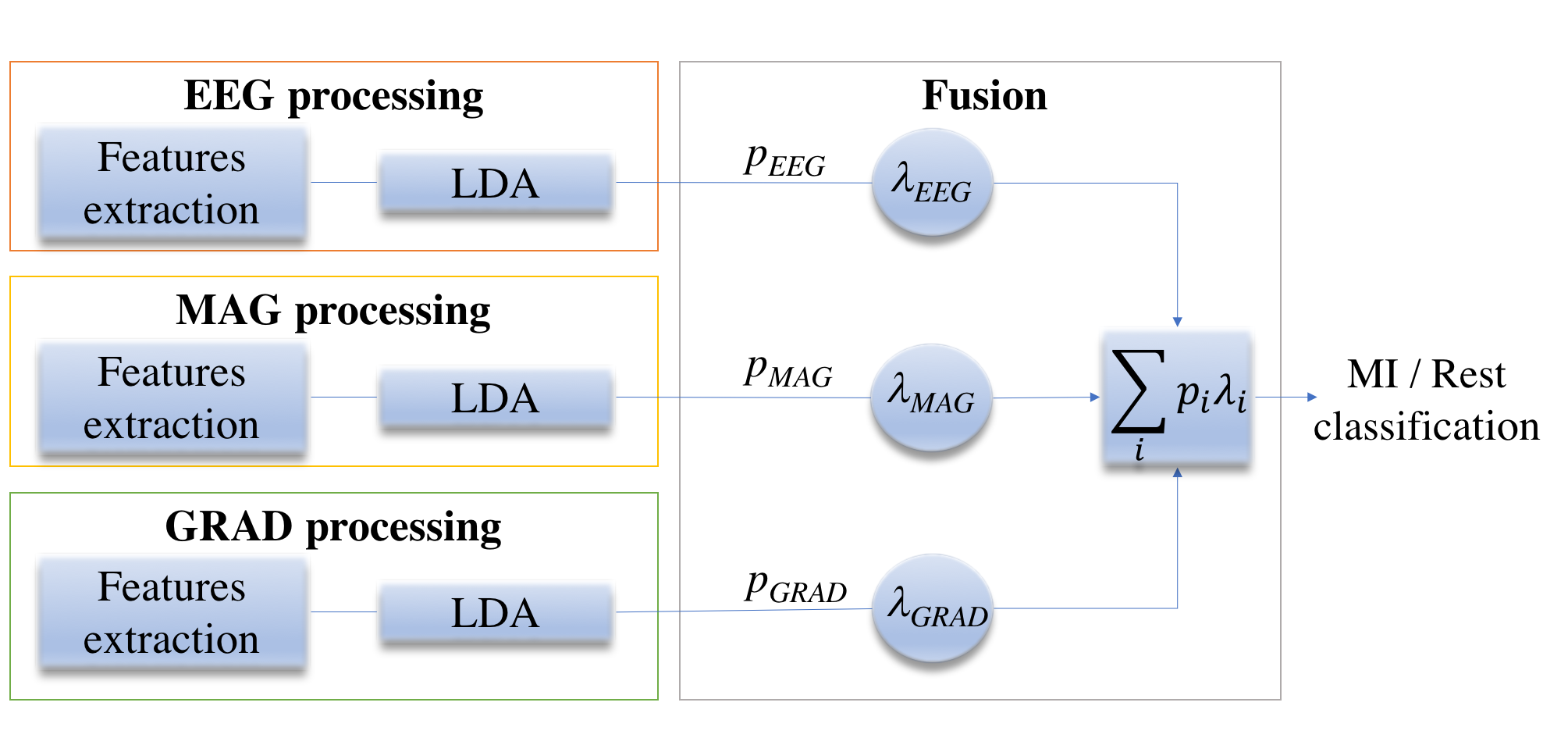} 
\caption{Classifier fusion approach for a given frequency bin. The variables $p_i$~and $~\lambda_i~$ stand for the posterior probability and the weight parameter associated with the modality i, respectively.}
\label{fig1}
\end{center}
\end{figurehere}

To assess the classifier performance, we measured the area under the receiver operating characteristic (ROC) curve (AUC) of the computed values of the false positive rate versus the true positive rate. AUC values typically range between one-half (chance level) and one (perfect classification) \cite{witten_data_2016}.
We evaluated our fusion approach with respect to the results obtained in each single modality separately (EEG, MAG, GRAD). In addition, we tested the effect of including an increasing number of most significant features. 

To statistically compare the results, we input the corresponding AUC values into a nonparametric permutation-based ANOVA with two factors: modality (EEG, MAG, GRAD, Fusion) and features ($N_{f}~=~1,2,...,10$). A statistical threshold of $p<0.05$ and 5000~permutations were fixed. We finally used a Tukey-Kramer method \cite{zar_biostatistical_1999} to perform a post-hoc analysis with a statistical threshold of $p<0.05$. This analysis was performed using routines available in the standard MATLAB and the EEGLAB toolboxes.  \cite{delorme_eeglab:_2004}

\section{Results}
Figure \ref{fig2} shows the grand-average time-frequency maps of the event-related de/synchronization (ERD/S) computed from the MI trials in the testing phase: \cite{pfurtscheller_event-related_1999}
$ERD/S = 100\times{\displaystyle x_{target} - \mu_{base}  \over \mu_{base}}$
where $x_{target}$ was the time-frequency energy of a sensor's signal for the target period $(1-6)$s and $\mu_{base}$ was the corresponding time-averaged energy in the baseline $(0-1)$s.

\begin{figurehere}
\begin{center}
\includegraphics[scale=0.33]{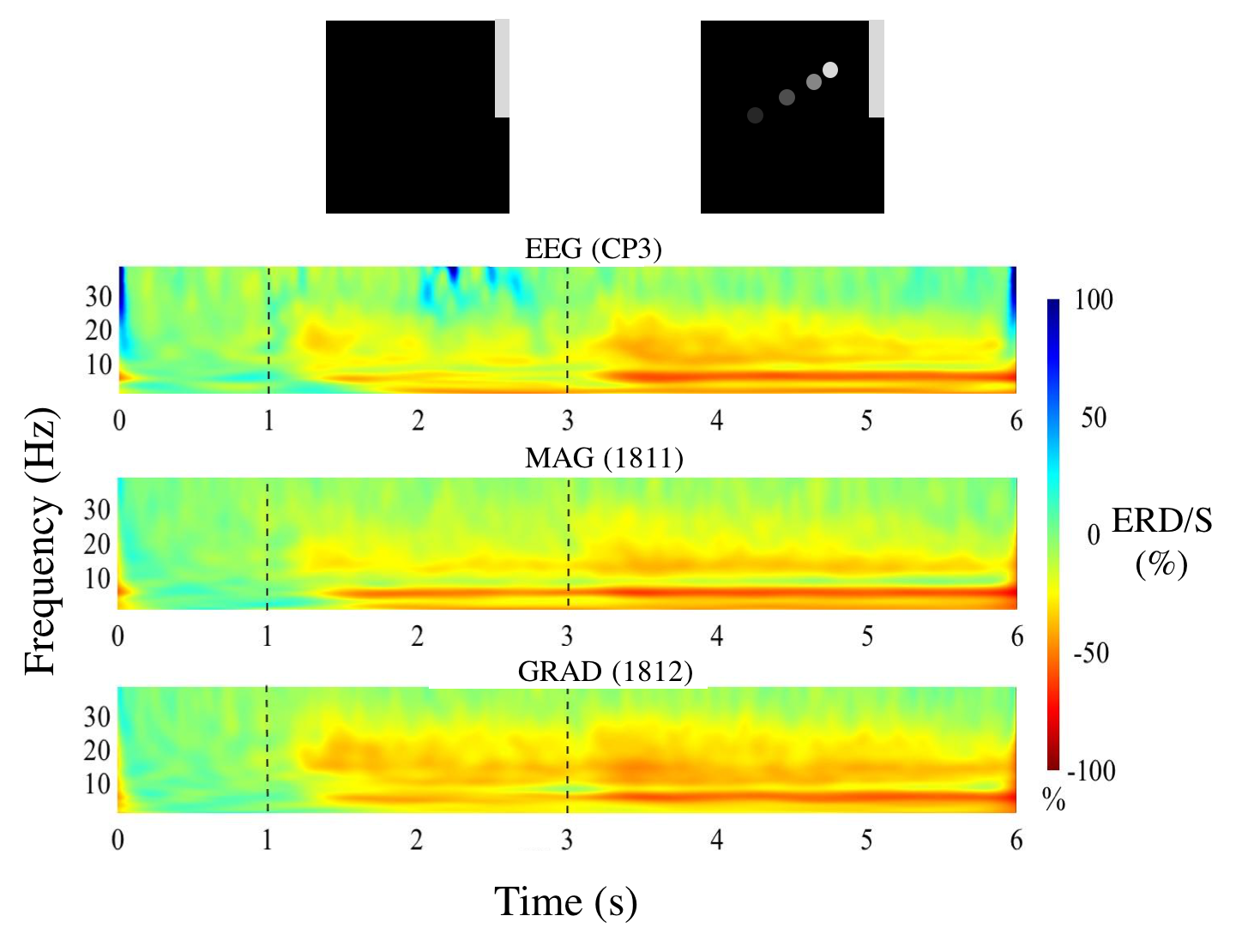}
\caption{Grand-average time-frequency maps of ERD/S. Top panels illustrate the visual stimulus  that appeared during the target period. Dashed lines mark the start of the target presentation and the feedback periods (see section 2.2, testing phase). 
The time-frequency decomposition of the signals was obtained through Morlet wavelets between 4~and 40~Hz, with a central frequency of 1~Hz associated with a time resolution of 3~s via the Brainstorm toolbox. \cite{tadel_brainstorm:_2011} Positive ERD/S values indicate percentage increases (i.e. neural activity synchronization), while negative values stand for percentage decreases (i.e. neural activity desynchronization)}.
\label{fig2}
\end{center}
\end{figurehere}
\noindent In all modalities, we observed significant changes for the alpha (ERD $\simeq -100\%$) and beta band (ERD $\simeq -60\%$). \cite{neuper_event-related_2001,neuper_imagery_2005,pfurtscheller_mu_2006} ERDs started to appear just after the target appearance (t~=~1~s) and became stronger during the feedback period (t~=~3-6~s). Notably, ERDs tended to appear early in the MEG signals as compared to the EEG signals. \\ 
\\
Figure 3~illustrates the candidate features that were selected through the semi-automatic procedure for each modality in the training phase. 
Features obtained from MEG signals tended to be more focused both in space (around the primary motor areas of the hand) and in frequency (mostly in the alpha band). This finding was in line with the fact that lower ERDs were observed in the beta band (see Figure \ref{fig2}).\\ 
\\

\begin{figurehere}
\begin{center}
\includegraphics[scale=0.2]{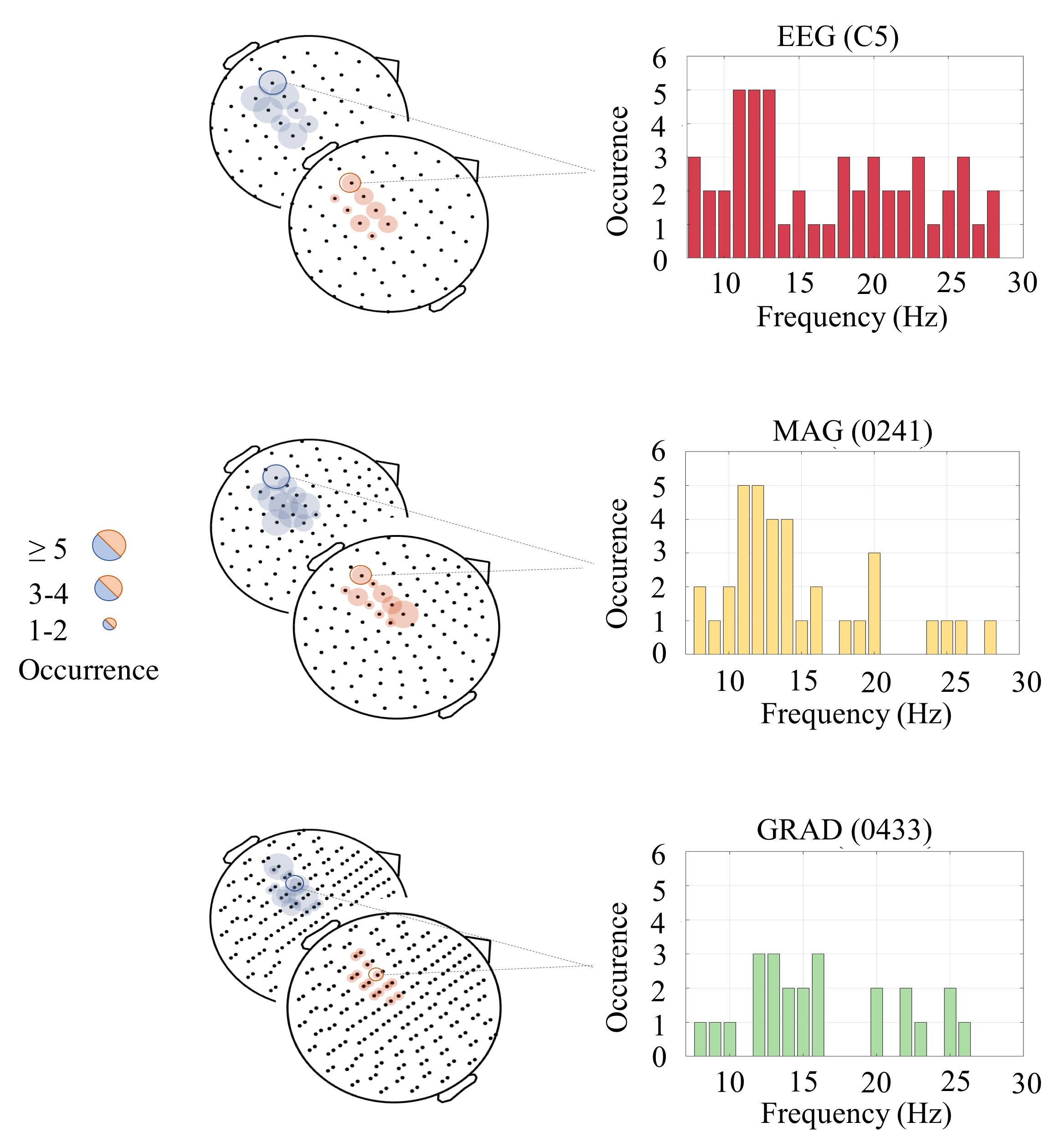} 
\caption{Spatial and frequency distribution of the features selected for classification in each modality. On the left side, the color of the nodes identifies the frequency band (blue for alpha band and red for beta band). The size of the circles is proportional to the number of subjects exhibiting that specific sensor as the best feature. On the right, the histograms detail the occurrences in every frequency bin for the sensor that was most frequently selected.}
\end{center}
\label{fig3}
\end{figurehere}

\noindent {\it Fusion improves classification performance}\\
In all frequency bands, the type of modality significantly affected the AUC values (ANOVA, $ p~<~10 ^{-3}$), whereas the number of features did not have a significant impact ($p~>~0.05$). The AUC values obtained with the fusion approach were significantly higher than those obtained with any other modality (Tukey-Kramer post-hoc, $p < 0.016$), except for theta and gamma bands for which we did not observe significant improvements with respect to EEG. The highest classification performance was obtained in the alpha band (Figure \ref{fig4}), for which we also reported here a significant interaction effect between modality and number of features (ANOVA, p~=~0.0069).  In this case, the AUC values with the fusion were significantly higher than those obtained with EEG, MAG, or GRAD separately (Tukey-Kramer post-hoc, $ p=4.3\times10^{-9},3.9\times10^{-7}$, and $0.012$). \\
To evaluate the classification performance in every subject, we considered for each modality the optimal number of features $N_{f}$ and the best frequency band associated with the highest AUC. 
Results showed that in thirteen subjects, the fusion led to a better performance as compared to single modalities, with AUC values ranging from $0.55$ to $0.85$, and relative increments ranging from $1.3\%$ to $50.9\%$ (with an average of $12.8\pm6\%$). 
In only three subjects, the fusion gave equivalent performance (see Table \ref{tab1}). 
More specifically, if we compared the performances obtained with the fusion with those resulting from the best single modality, the average improvement was of $4\pm3\%$. Noteworthy, this value was of $15 \pm 17\%$ when we compared the fusion with EEG, the modality that is the most used during BCI experiments. \cite{clerc_brain-computer_2016,clerc_brain-computer_2016-1} 

Interestingly, the contribution of the different modalities to the fusion\textquotesingle s performance was highly variable across subjects, as illustrated by the weights associated to the parameter $\lambda_i$ (see Figure \ref{fig5}).

\begin{figurehere}
\begin{center}
\includegraphics[scale=0.33]{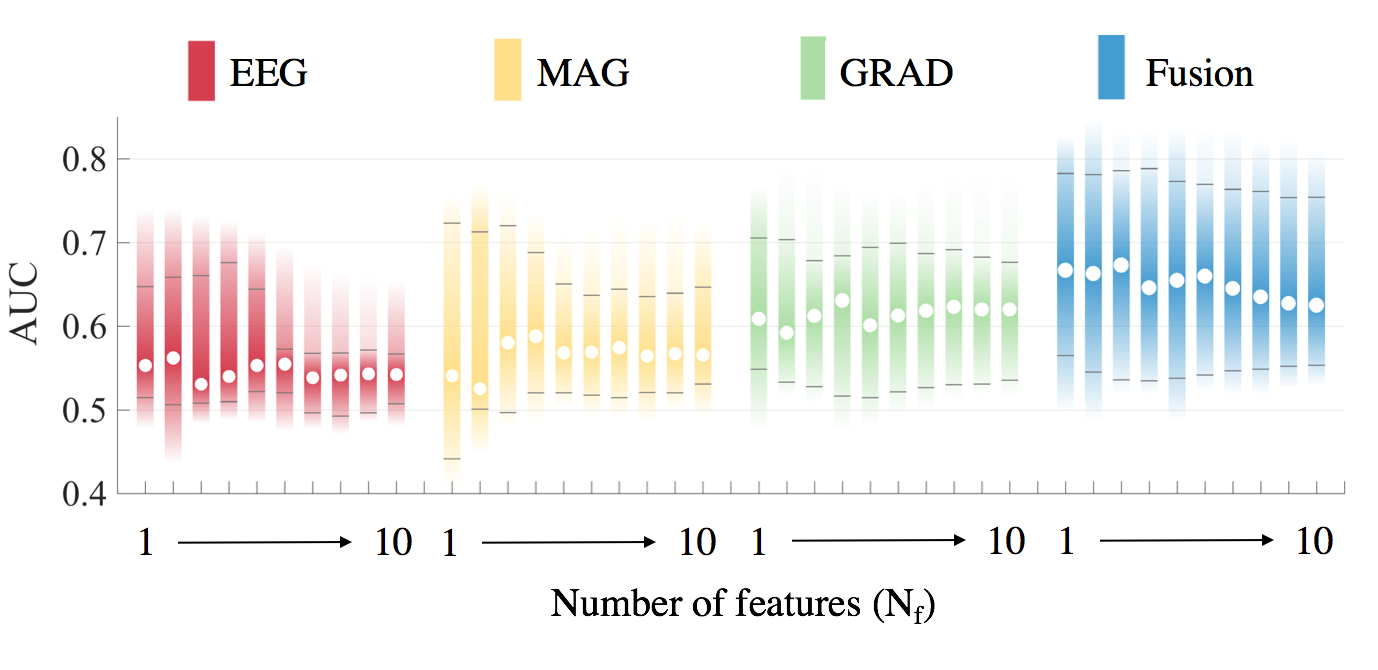}
\caption{AUC distributions across the 15 subjects, for different modalities (EEG, MAG, GRAD, Fusion) and for different number of features $N_f$ in the alpha band. White circles represent the associated median values.}
\label{fig4}
\end{center}
\end{figurehere}

\end{multicols}
\begin{table}
\tbl{Individual performances overview across modalities. In bold, the best AUC obtained for a given subject. 
\label{tab1}}
{\fontsize{7}{11}\selectfont
\begin{tabular}{@{}cccccccccccccccc@{}}
\toprule
 & S01 & S02 & S03 & S04 & S05 & S06 & S07 & S08 & S09 & S10 & S11 & S12 & S13 & S14 & S15 \\ 
  \botrule
 {\it Band} & {\it alpha} & {\it alpha} & {\it beta} & {\it alpha} & {\it alpha} & {\it alpha} & {\it beta} & {\it beta} & {\it alpha} & {\it alpha} & {\it alpha} & {\it alpha} & {\it beta} & {\it alpha} & {\it beta} \\
 \botrule
EEG & 0.53 & \textbf{0.55} & 0.48 & 0.56 & 0.57 & 0.55 & \textbf{0.57} & 0.50 & 0.60 & 0.66 & 0.53 & 0.55 & 0.70 & 0.73 & 0.60 \\ 
MAG & 0.47 & 0.48 & 0.53 & 0.51 & 0.50 & 0.55 & 0.52 & 0.62 & 0.57 & 0.58 & 0.69 & 0.76 & 0.59 & 0.64 & 0.71 \\ 
GRAD & 0.54 & 0.50 & 0.55 & 0.50 & 0.54 & 0.55 & 0.49 & 0.65 & 0.64 & 0.63 & 0.72 & 0.62 & 0.71 & 0.72 & \textbf{0.85} \\ 
Fusion & \textbf{0.55} & \textbf{0.55} & \textbf{0.56} & \textbf{0.57} & \textbf{0.58} & \textbf{0.59} & \textbf{0.57} & \textbf{0.67} & \textbf{0.66} & \textbf{0.70} & \textbf{0.80} & \textbf{0.77} & \textbf{0.76} & \textbf{0.79} & \textbf{0.85} \\ 
\botrule
\end{tabular}}
\end{table}
\begin{multicols}{2}

\section{Discussion} 
Improving performance remains one of the most challenging issues of non-invasive BCI systems.\cite{lotte_review_2007} High classification performance would allow effective control of the BCI and feedback to the subject that is crucial to establish an optimal interaction user-machine. \cite{clerc_brain-computer_2016, clerc_brain-computer_2016-1,vidaurre_towards_2010} BCI performance depends on several human and technological factors, including the ability of subjects to generate distinguishable brain features \cite{wolpaw_braincomputer_2002}, as well as the robustness of signal processing and classification algorithms. \cite{lotte_review_2007}

To this end, we recorded simultaneous EEG and MEG signals in a group of healthy subjects performing a motor imagery-based BCI task. Both EEG and MEG exhibit a high temporal resolution and the sensory motor-related changes are well known in the literature, as is their utility in standard BCI applications. \cite{mellinger_meg-based_2007,vidaurre_towards_2010} 

Notably, EEG and MEG signals are closely related but still they are respectively different in terms of sensitivity to radial and tangential currents, as well as to extracellular and intracellular currents \cite{hamalainen_magnetoencephalography-theory_1993}. These complementary properties could be simultaneously exploited by our fusion approach to better identify ERD mechanisms used here to control the BCI.
 
Results show that independently from the modality and the number of features, the best AUCs were obtained in alpha and (in a more limited way) beta bands, which is consistent with motor imagery\textquotesingle s being associated with oscillations in the alpha and beta band.  \cite{pfurtscheller_eeg-based_1997,pfurtscheller_motor_2001} 
The proposed fusion approach showed that combining the most significant features in each modality led, in a large majority of subjects, to a reduction in the subjects\textquotesingle~mental state misclassifications (see Table \ref{tab1}). By optimizing the choice of the features in each individual, we obtained an average classification improvement of 12.8~\% as compared to separate EEG, MAG and GRAD classifiers (Table \ref{tab1}), suggesting a viable alternative to indirectly reduce the illiteracy phenomenon in non-invasive BCIs. \cite{mcfarland_brain-computer_2011,erp_brain-computer_2012} 

In this study, we also explored features from other frequency bands such as the gamma band (30-40 Hz). However, the obtained results gave marginal improvements as compared to alpha and beta bands. 
While gamma activity from intracranial recordings or local-field potentials is in general related to the initiation of motor/sensory function \cite{tallon-baudry_oscillatory_1999, rodriguez_perceptions_1999, canolty_high_2006, cheyne_self-paced_2008, muthukumaraswamy_functional_2010, wander_distributed_2013}, the paucity of results in the gamma band could be here partly explained by the low signal-to-noise ratios and volume conduction effects that typically affect scalp EEG and MEG activity.  \cite{grosse-wentrup_causal_2011,grosse-wentrup_high_2012,jeunet_predicting_2015}

The core of our approach consisted in weighting automatically the contribution of each modality in an effort to optimize performance. This is an important aspect as the discriminant power of features could suddenly change depending on many factors, such as impedance fluctuations or the presence of artifacts (e.g. isolated EEG electrode pops or MEG jumps). In this case, our fusion approach would take into account such transient fluctuations by silencing the affected modality through a lower weight $\lambda_i$ in the classification.
Slower changes could be related to the increasing ability of individuals to accurately control the BCI. \cite{lotte_review_2007, curran_learning_2003, dobkin_braincomputer_2007, grosse-wentrup_using_2011,clerc_brain-computer_2016} In this case, our approach would progressively favor the spatio-temporal features of the modality that better capture those neural plasticity phenomena. 

Interestingly, we noticed a high inter-subject variability in the attributed weights (Figure \ref{fig5}). While, this could be associated with the ability of each modality to detect different properties of the underlying ERD, further analysis, possibly in the source space, \cite{gross_good_2013} is needed to elucidate this aspect and identify the neurophysiological correlates of such variability.

While the average AUC values were relatively low, we noticed that they are highly variable across individuals (Table \ref{tab1}) and that they are close to those typically obtained in similar experiment settings. \cite{pfurtscheller_hybrid_2010} Furthermore, it is important to mention that subjects were BCI-na\"ive and that no preprocessing was applied, with the goal of simulating real-life scenarios. Thus, while a proper pre-processing was likely.

\begin{figurehere}
\begin{center}
\includegraphics[trim={2.75cm 0 0 0}, scale=0.42]{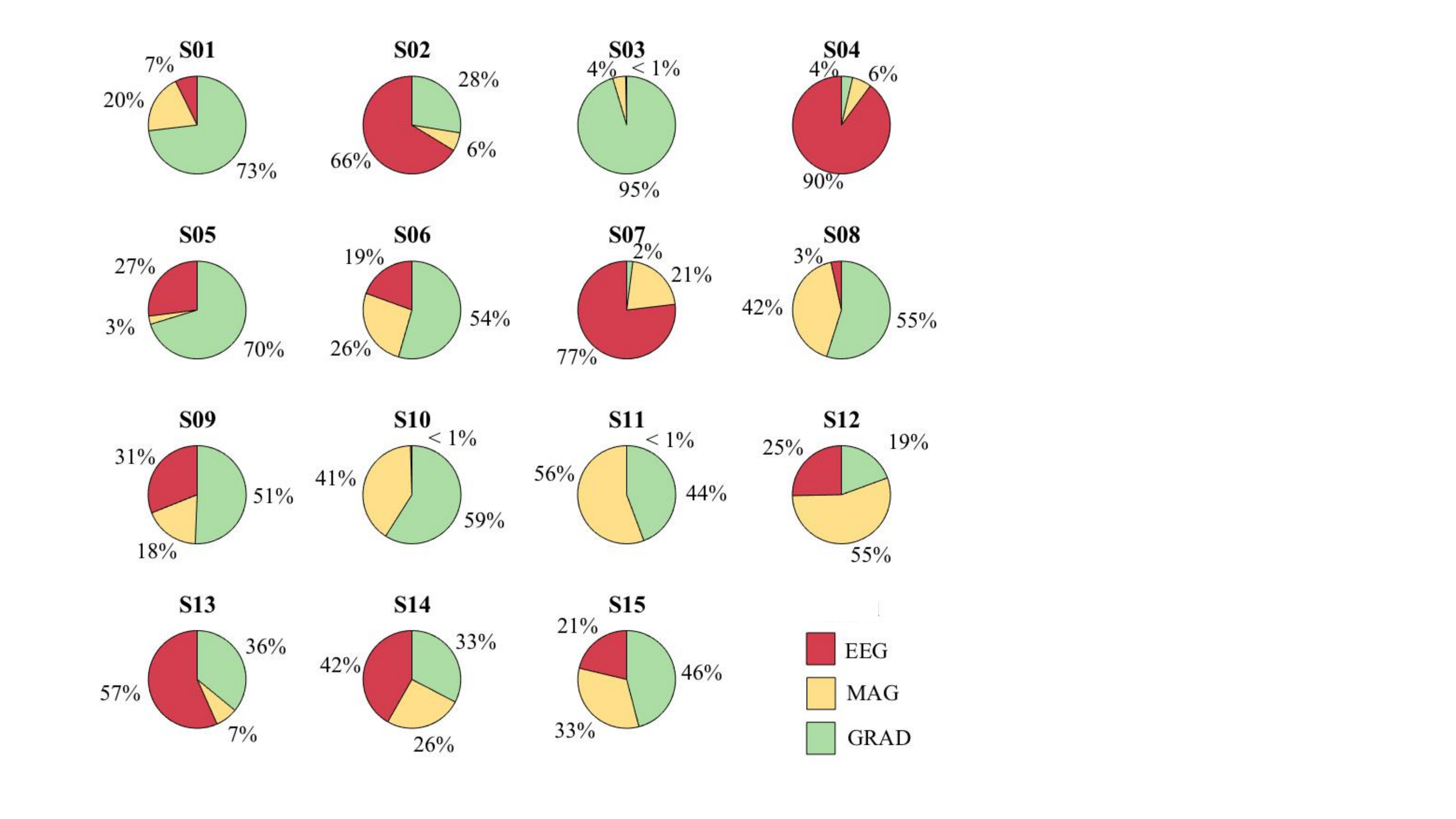}
\caption{Contribution of different modalities to the individual performance. Pie-diagrams show the $\lambda_i$ values (in percentage) obtained for each modality via the fusion approach.}
\label{fig5}
\end{center}
\end{figurehere}

\noindent to improve the accuracy in each single modality, our aim was rather to assess  an improvement in the worst condition. Eventually, thirteen of our fifteen subjects presented a performance improvement with the classifier fusion.

Taken together, these results prove the potential advantage of using simultaneous E/MEG signals to enhance BCI performance. By using a rather simple classifier (LDA), we could include in the classification a reduced number of specific features involved in the motor-related neural mechanisms such as ERD in alpha and beta bands. 

More sophisticated approaches using the whole feature space, such as support vector machines \cite{lai_support_2004} and Riemannian geometry \cite{barachant_classification_2013}, as well as alternative fusion strategies, such as boosting, voting, or stacking strategies \cite{lotte_review_2007}, but also classification in source space to improve spatial resolution \cite{sharon_advantage_2007,muthuraman_beamformer_2014} and identification of subject-specific time-frequency characteristics \cite{yang_time-frequency_2014}, can be further evaluated to exploit their power in practical applications. 

Finally, it is important to note that we tested our fusion approach offline by analyzing previously recorded data.
To evaluate the feasibility in online applications, we estimated that for an epoch of $500$ ms the time necessary to compute the features, perform the classification, and determine the parameter of the fusion, was approximately of $20$ ms when $N{_f}$~=~5. This value is actually compatible with current on-line settings using similar time windows and updating the feedback every $28$ ms. \cite{schalk_bci2000:_2004}

\section{Conclusions}
Our results showed that integrating information from simultaneous EEG and MEG signals improves BCI performance. E/MEG multimodal BCIs may turn out to be an effective approach to enhance the reliability of brain-machine interactions, but much of the progress will depend on the miniaturization of MEG scanners, which currently require a magnetic shielding room (MSR) and sensors cooled via a cryogenic system. 
Recent efforts proposing miniaturized and cryogenic-free MEG sensors \cite{jimenez-martinez_microfabricated_2017, boto_new_2017} and avoiding the use of MSRs \cite{sorbo_unshielded_2017} will hopefully offer practical solutions to increase MEG portability and boost the development of multimodal BCIs.

\section{Acknowledgements} 
We would like to thank the anonymous reviewers for their constructive comments and suggestions. This work was partially supported by French program ``Investissements d\textquotesingle avenir" ANR-10-IAIHU-06; ``ANR-NIH CRCNS" ANR-15-NEUC-0006-02 and by Army Research Office (W911NF-14-1-0679). The funders had no role in study design, data collection and analysis, decision to publish, or preparation of the manuscript.

\renewcommand{\bibname}{References}
\bibliographystyle{ws-ijns} 
\bibliography{FusionArticle}
\end{multicols}

\newpage
\appendix{}
\begin{figurehere}
\begin{center}
\includegraphics[scale=0.6]{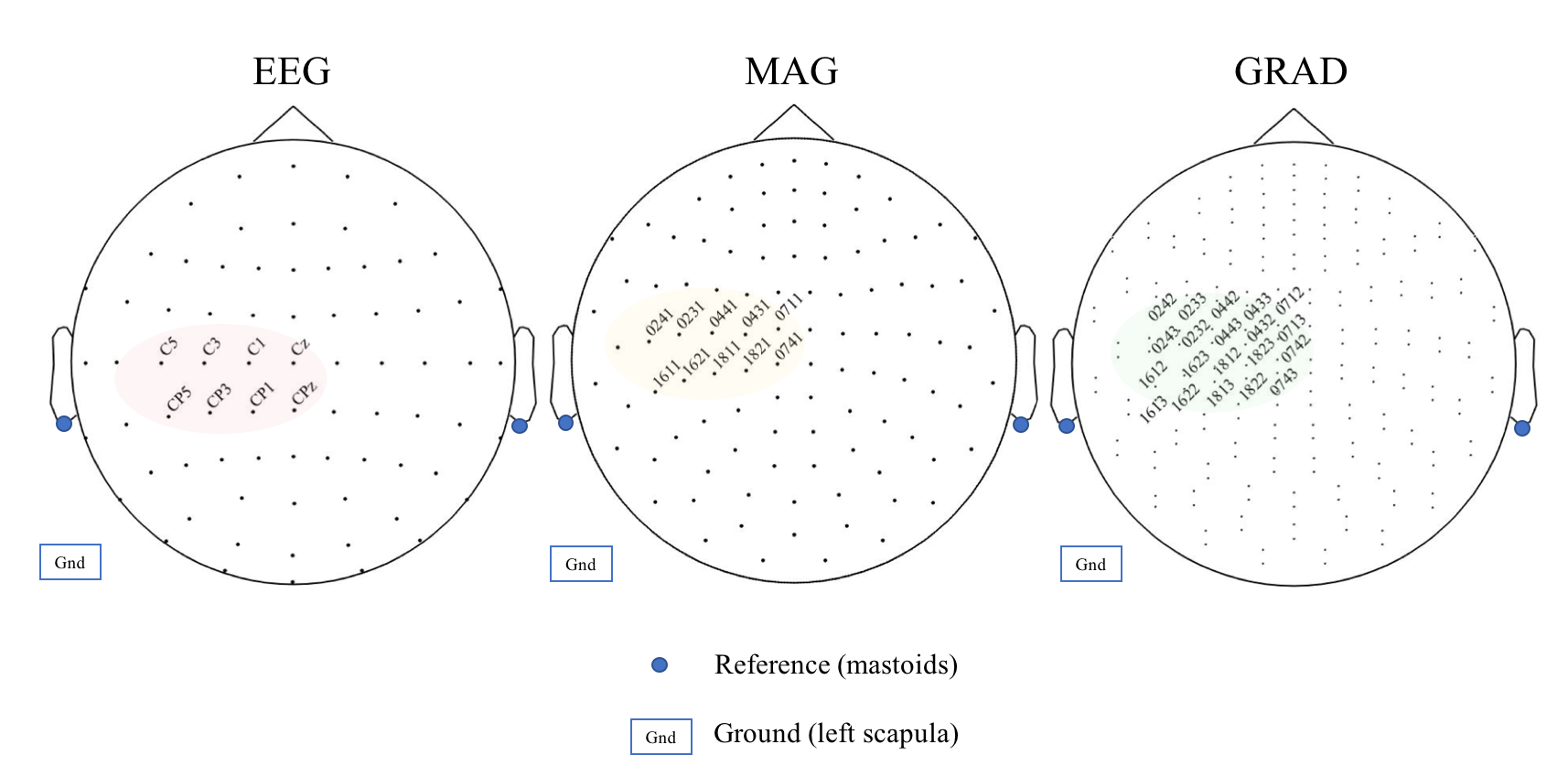}
\caption{Pre-selected EEG and MEG sensors (left motor area).}
\label{appendix1}
\end{center}
\end{figurehere}

\end{document}